\documentclass[%
 aip,
 amsmath,amssymb,
 reprint,%
]{revtex4-1}

\usepackage{graphicx}
\usepackage{dcolumn}
\usepackage{bm}

\usepackage[utf8]{inputenc}
\usepackage[T1]{fontenc}
\usepackage{mathptmx}
\usepackage[dvipsnames]{xcolor}

\newcommand{\Ka}[0]{{$K{\alpha} $}}
\newcommand{\Kah}[0]{{$K{\alpha}^{h} $}}

\newcommand{\feo}[0]{$\gamma-$Fe$_2$O$_3$}
\newcommand{\La}[0]{{$L{\alpha} $}}

\begin{document}

\preprint{AIP/123-QED}

\title{Fluorescence intensity correlation imaging with high resolution and elemental contrast using intense x-ray pulses}

\author{Phay J. Ho}
\email{pho@anl.gov}
\affiliation{Chemical Sciences and Engineering Division, Argonne National Laboratory, Lemont, Illinois 60439, USA}

\author{Christopher Knight}
\affiliation{Computational Science Division, Argonne National Laboratory, Argonne, Illinois 60439, USA}

\author{Linda Young}
\affiliation{Chemical Sciences and Engineering Division, Argonne National Laboratory, Argonne, Illinois 60439, USA}
\affiliation{Department of Physics and James Franck Institute, The University of Chicago, Chicago, Illinois 60637, USA}

\begin{abstract}

We theoretically investigate the fluorescence intensity correlation (FIC) of Ar clusters and Mo-doped iron oxide nanoparticles subjected to intense, femtosecond and sub-femtosecond XFEL pulses for high-resolution and elemental contrast imaging.  We present the FIC of {\Ka} and {\Kah} emission in Ar clusters and discuss the impact of sample damage on retrieving high-resolution structural information and compare the obtained structural information with those from the coherent difractive imaging (CDI) approach.  We found that, while sub-femtosecond pulses will substantially benefit the CDI approach, few-femtosecond pulses may be sufficient for achieving high-resolution information with FIC.  Furthermore, we show that the fluorescence intensity correlation computed from the fluorescence of Mo atoms in Mo-doped iron oxide nanoparticles can be used to image dopant distributions.


\end{abstract}

\maketitle


\section{Introduction}

Coherent diffractive imaging (CDI) \cite{Neutze-2000-Nature,Chapman-2006-NatPhys} with x-ray free-electron laser (XFEL) pulses holds the promise to probe the structure \cite{Seibert-2011-Nature,Ekeberg-2015-PRL} and follow the dynamics \cite{Clark-2013-Science, Gorkhover-2016-NatPho, Ferguson-2016-ScienceAdv} of entities with atomic resolution \cite{Neutze-2000-Nature}.  
This approach is based on the “probe before destroy" concept and makes use of the very high number of photons in a single pulse.
The idea is that with short-duration pulses, high-resolution diffraction patterns can be captured with a single XFEL pulse before the system of interest suffers damage from the intense radiation \cite{Neutze-2000-Nature}.  Despite continuous progress, it has remained a challenge \cite{Aquila-2015-StrucDyn, Ho-2020-NatComm} to achieve nanometer or subnanometer resolution and elemental contrast \cite{Gomez-2014-Science} with CDI.  This is because intense x-ray pulses will lead to extremely rapid structural degradation of the sample and generate a large number of delocalized electrons, resulting in a substantial reduction in both scattering efficiency \cite{Ho-2016-PRA, Ho-2020-NatComm} and signal-to-noise in the measured scattering patterns  \cite{Quiney-2011-NatPhy,Lorenz-2012-PRE,Ziaja-2012-NJP}. 

To circumvent the challenges of CDI approach, new XFEL imaging modalities are being developed for Single Particle Imaging (SPI).  Recently, an approach using single-shot fluorescence intensity correlation (FIC) has been proposed to measure the 3-D distribution of heavy elements using intense x-ray pulses \cite{Classen-2017-PRL}.  This approach was inspired by the Hanbury Brown and Twiss (HBT) effect \cite{Hanbury-1956-Nature}, which has enabled the determination of the size of astronomical objects from the intensity correlation of their emitted light.  
The discovery of the HBT effect has pushed the imaging resolution below the Abbe limit (classical diffraction limit) in the visible range \cite{Thiel-2007-PRL, Dertinger-2009-PNAS, Monticone-PRL-2014, Yu-2016-PRL,Oppel-2014-PRL, Pelliccia-2016-PRL,Classen-2016-PRL,Oppel-2014-PRL}, as well as in the XUV regime \cite{Schneider-2017-NatPhys}.  
Also, recent experiments demonstrated that FIC can be used to directly determine the structure of trapped ion pairs \cite{Richter-2020-arxiv,Wolf-2020-PRL}. 
We point out that an alternative fluorescence imaging approach has been proposed by Ma and coworkers to measure the molecular structure from the interference of the emitted fluorescence from x-ray excited molecules with identical atoms \cite{Ma-1994-CPL,Ma-1995-RSI}.  Unlike FIC, this approach is related to measuring the first-order field correlation functions.

\begin{figure*}[!t]
  \begin{center}
\includegraphics[width=6.5 in]{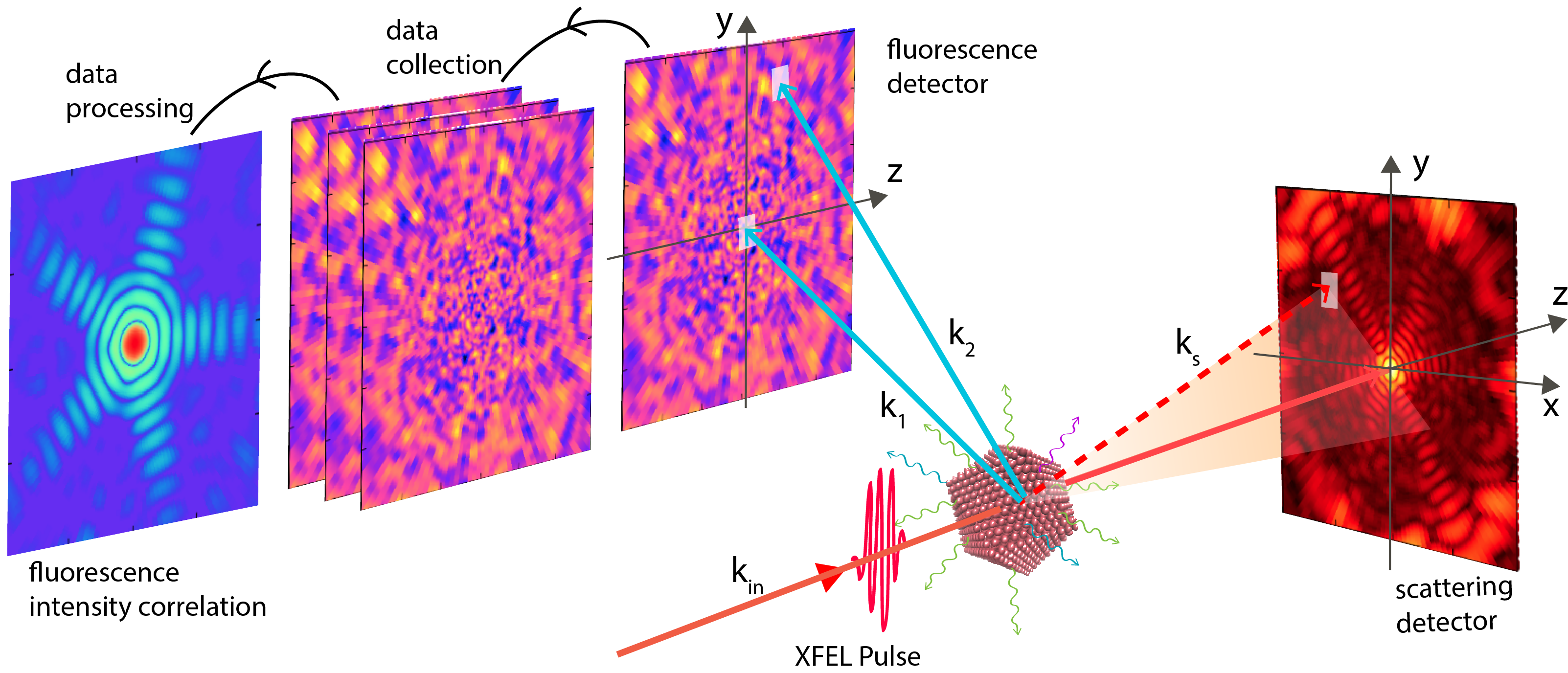}
\caption{Setup of SPI with coincident fluorescence and scattering imaging. Fluorescence detector is placed perpendicular to the x-ray beam, where the coherent scattering is suppressed.  In principle, it can be placed independent of the beam since fluorescence emission is isotropic.  In FIC, the structural information is obtained from the correlation of many pairs of spatially separated pixels.}
\label{fig:setup}
  \end{center}
\vspace{-15pt}
\end{figure*}

With the ever-improving XFEL technology and the emerging new operation modes that deliver intense few- and sub-fs pulses at high repetition rates \cite{Duris-2020-NatPho}, the XFEL will further accommodate this photon-hungry fluorescence intensity correlation measurement \cite{Schneider-2017-NatPhys} in both the soft and hard x-ray regimes.  In the past few years, several research groups have begun experimental and theoretical studies of fluorescence imaging that employs XFEL pulses.   For example, FIC method has been used to determine both the pulse duration \cite{Inoue-2019-JSR} and focal area \cite{Nakamura-2020-JSR} of femtosecond XFEL pulses.  Recent simulation work by Trost and coworkers suggests that while it is challenging to image macroscopic objects due to low signal-to-noise ratios, FIC may be useful for imaging of single particles \cite{Trost-2020-NJP}.  Our previous study, which used Ar clusters as a prototypical system, showed that the intense, few-femtosecond XFEL pulses can enable a multitude of atomic fluorescence channels, but their dynamics in an extended system is drastically different from that in isolated atoms.  We found that the x-ray emission time in extended systems is dictated by not only the x-ray pulse duration and lifetime of the core-excited states. Rather, ultrafast electron dynamics can greatly modify fluorescence characteristics by opening additional fluorescence pathways via electron-ion recombination in addition to the direct photoionization pathways.  The existence of these two pathways gives rise to higher {\Ka} and {\Kah} yields and broader temporal emission profiles in clusters.   

Building on the results of our previous work, we theoretically explore FICs for high-resolution and elemental contrast imaging of isolated nanosized systems with intense x-ray pulses.  
We investigate the FIC of {\Ka} and {\Kah} emission in Ar clusters as a function of pulse duration and discuss the impact of sample damage on FIC.  We show that FIC is less sensitive to radiation damage than CDI and can be exploited for retrieving high-resolution structural information.  Furthermore, we examined the FICs computed from the fluorescence of Mo atoms in Mo-doped iron oxide hollow core shell nanoparticles (NPs). In these particles, the macroscopic properties and catalytic functions are sensitive to the presence of Mo dopants and the structure of the NPs \cite{Kwon-2016-NL}.  However, the influence of the dopant is not yet known as the atomic-scale structures and, in particular, the dopant distributions in the shell are not known \cite{Kwon-2016-NL,Kwon-2015-NatMat}.   Our calculations show that FIC is more sensitive to the dopant distribution than CDI. 

\section{Model}

To model the FIC of an extended system, we treat individual fluorescent atoms as random light emitters, such that the intensity of the combined emitted radiation field at $\Vec{k}$ can be expressed as
\begin{equation}
    I(\Vec{k}) = I_0 \sum_{j=1}^{N_{emitters}}\sum_{l=1}^{N_{emitters}}\beta_j \beta_l^* e^{i\Vec{k} \cdot (\Vec{r}-\Vec{R}_{f,j})} e^{-i\Vec{k} \cdot (\Vec{r}-\Vec{R}_{f,l})}e^{i (\phi_j-\phi_l)}
\end{equation}
where $\Vec{R}_{f,j}$ is the position and $\phi_j$ the random phase of the $j$-th emitter.  Here, $\phi_j$ is related to the emission time of the random fluorescence event. $I_0=\langle I(\Vec{k},\{\Vec{R}_{f,j}\}) \rangle$ , which is is independent of $\Vec{k}$, is the average intensity obtained by averaging over many realizations of the random fluorescence events, and $\beta_j$ is considered as the fractional emitter strength such that
\begin{equation}
    \sum_{j}^{N_{emitter}} |\beta_j|^2 = 1.
\end{equation}
For a collection of noninteracting emitters (i.e. in gas phase), $|\beta_j|^2=1/N_{emitter}$.  But, for an extended and condensed system in an intense x-ray pulse, this is not true and $\beta_j$ can depend on the location of the emitter.  This is because the x-ray induced ionization and the subsequent electron-ion recombination and massive electron rearrangement can quickly transform the exposed sample volume into a nonuniform spatial charge density structure (neutral core and highly charged shell)\cite{Rupp-2016-PRL,Ho-2017-JPB}.



In this model, the FIC is the second-order correlation function of the fluorescence intensity measured at two momentum transfer vectors, $I(\Vec{k}_1$) and $I(\Vec{k}_2)$, and is related to the Fourier transform of the distribution of the fluorescence emitters $\rho_f$ by the following relation \cite{Baym-1998-arXiv,AGARWAL202013}
\begin{equation}
    G{_2}(\Vec{k}_1,\Vec{k}_2,\{\Vec{R}_{f,j}\}) - 1 = \bigg\lvert \int \!\! d^3r \rho_f(\vec{r},\{\Vec{R}_{f,j}\}) e^{i\Vec{q}_f\cdot \Vec{r}} \bigg\rvert^2
\end{equation}
where $\Vec{q}_f=\Vec{k}_1-\Vec{k}_2$ is the momentum transfer vector and $\{\Vec{R}_{f,j}\}$ denotes the spatial configuration of all the emitters and
\begin{equation}
    G{_2}(\Vec{k}_1,\Vec{k}_2,\{\Vec{R}_{f,j}\}) = \frac{\langle I(\Vec{k}_1,\{\Vec{R}_{f,j}\})I(\Vec{k}_2,\{\Vec{R}_{f,j}\}) \rangle}{\langle I(\Vec{k}_1,\{\Vec{R}_{f,j}\}) \rangle \langle I(\Vec{k}_2\{\Vec{R}_{f,j}\}) \rangle},
\end{equation}
\begin{equation}
\rho_f(\vec{r},\{\Vec{R}_{f,j}\}) =  \sum_{j}^{N_{emitter}} |\beta_j|^2  \delta(\vec{r}-\Vec{R}_{f,j})
\end{equation}
An intense x-ray pulse can strongly excite the sample leading to the changes in the spatial arrangement of the emitters during the time window of fluorescence.  The measured $G_2$ needs to be weighted over all the spatial configurations.
\begin{equation}
    \langle G{_2}(\Vec{k}_1,\Vec{k}_2,\{\Vec{R}_{f,j}\}) \rangle_{avg} = \sum_{\{\Vec{R}_{f,j}\}} p(\{\Vec{R}_j\}) G{_2}(\Vec{k}_1,\Vec{k}_2,\{\Vec{R}_{f,j}\})
\end{equation}
where $p(\{\Vec{R}_{f,j}\})$ is the probability of the collection of emitters in the configuration of $\{\Vec{R}_{f,j}\}$ and $\sum_{\{\Vec{R}_{f,j}\}} p(\{\Vec{R}_{f,j}\})= 1$.  

To determine $G_2$, we used our previously developed Monte Carlo/Molecular Dynamics (MC/MD) method.  This is an effective method for describing matter interacting with intense x-ray pulses and it has been used to reproduce XFEL experimental data: ion kinetic energy distribution of Ar clusters \cite{Ho-2017-JPB}, ion charge state distribution of Ar atoms \cite{Ho-2014-PRL} and the ultrafast x-ray scattering response of molecular clusters \cite{Ho-2020-NatComm}.   The details of this method are described in our previous work \cite{Ho-2017-JPB,Ho-2016-PRA}.   In brief, the interaction of the cluster of atoms with the incident XFEL pulse is treated quantum mechanically with a Monte Carlo method by tracking explicitly the time-dependent quantum transition probability between different electronic configurations. The total transition rate, $\Gamma$, between different electronic configurations $I$ and $J$ is given by
\begin{equation}
\Gamma_{I,J} = \Gamma^{P}_{I,J}+\Gamma^{A}_{I,J}+\Gamma^{F}_{I,J}+\Gamma^{RE}_{I,J}+\Gamma^{EI}_{I,J}+\Gamma^{RC}_{I,J}.
\end{equation}
Starting from the ground state of the neutral atom, we include the contribution from photoionization $\Gamma^{P}_{I,J}$, Auger decay $\Gamma^{A}_{I,J}$, fluorescence $\Gamma^{F}_{I,J}$, resonant excitation $\Gamma^{RE}_{I,J}$, electron-impact ionization $\Gamma^{EI}_{I,J}$, and electron-ion recombination $\Gamma^{RC}_{I,J}$.  Additionally, a molecular dynamics (MD) algorithm is used to propagate all particle trajectories (atoms/ions/electrons) forward in time.  By tracking the time evolution of electronic configurations of atoms and ions (timing of each fluorescence event) and their positions, MC/MD can study the impact of an intense x-ray pulse on various fluorescence channels, as well as the associated FICs, in one calculation.  Since our analysis focuses on the intense pulse effect on the fluorescence dynamics, we assume isotropic x-ray emission \cite{Southworth-1991-NIMPR}.

In this paper, we further exploit the capability of MC/MD model to investigate multiple imaging modalities.  In addition to FIC, we also examine the x-ray scattering pattern under the same pulse conditions.  The scattering response is modeled as a sum of the instantaneous scattering patterns weighted by the pulse intensity, $j_X(\tau,t)$ with FWHM duration $\tau$.  In our model, the scattering signals expressed in terms of the total differential scattering cross section of the target system can be regarded as the sum of the coherent (elastic) and incoherent (inelastic) scattering \cite{Hubbell-1975-JPCRD, Chihara-1987-JPF, Crowley-2014-HEDP2, Slowik-2014-NJP} 
\begin{eqnarray}
  \label{eq:totalSCS}
  \frac{d  \sigma_{total}}{d\Omega}(\Vec{q}) =\frac{d  \sigma_{coh}}{d\Omega}(\Vec{q}) + \frac{d  \sigma_{incoh}}{d\Omega}(\Vec{q}) \ ,
\label{eq:totalDCS}
\end{eqnarray}
where the coherent scattering can be expressed as
\begin{eqnarray}
  \label{eq:npSCS}
 \frac{d\sigma_{coh}}{d\Omega}(\Vec{q}) =  \frac{d\sigma_{KN}}{d\Omega} \frac{1}{F}\int_{-\infty}^{+\infty}\!\! dt  j_X(\tau,t) |F_{c}(\Vec{q},t)|^2 ,
\end{eqnarray}
where $d\sigma_{KN}/d\Omega$ is the Klein-Nishina scattering cross section \cite{Klein-1929-ZFP} and $F = \int_{-\infty}^{+\infty}\!\! dt  j_X(\tau, t)$ is the fluence of an XFEL pulse.  Here,
\begin{equation}
  \label{eq:timeFmol}
   F_{c}(\Vec{q},t) = \int d^{3} \Vec{r} \rho_{1e}(\Vec{r};\{\Vec{R}_{j}\},t) e^{i\Vec{q} \cdot\Vec{r}}.
\end{equation}
is the time-dependent form factor of the target system, where $\rho_{1e}(\Vec{r};\{\Vec{R}_{j}\},t)$ is the time-dependent electron density of the system with a geometry, $\{\Vec{R}_{j}\}$.  In this reference frame, as shown in Fig. \ref{fig:setup}, the momentum transfer vector is $\Vec{q} = \Vec{k}_{in} - \Vec{k}_{s}$, where $\Vec{k}_{in}$ and $\Vec{k}_{s}$ are the wave vectors of the incident and scattered photons.

By using the independent atom model (IAM), $F_{c}(\Vec{q},t)$ can be written as
\begin{equation}
  \label{eq:npFormFactor}
 F_{c}(\Vec{q},t)=\sum_{j=1}^{N_a} f_j(\Vec{q},C_j(t))e^{i \Vec{q} \cdot \Vec{R}_j(t)} + \sum_{j=1}^{N_e(t)} e^{i \Vec{q} \cdot \Vec{r}_j(t)}\, ,
\end{equation}
where $N_a$ is the total number of atoms/ions, $\Vec{R}_j(t)$, $C_j(t)$ and $f_j(\Vec{q},C_j(t))$ are the position, the electronic configuration and the atomic form factor of the $j$-th atom/ion respectively.  $N_e(t)$ is the number of delocalized electrons within the focal region of the x-ray pulse and $\Vec{r}_j(t)$ are their positions.  Our previous work shows that the IAM works well for describing the intense-pulse scattering response of molecular clusters \cite{Ho-2020-NatComm} and single molecules, in which the difference between the IAM and DFT methods, which go beyond IAM and include electron correlations, is of the order a few percents \cite{Ho-2021-FD}. 

The contribution from incoherent scattering processes is cast in terms of the incoherent scattering function, $ S(\Vec{q},t)$ \cite{Hubbell-1975-JPCRD}:
\begin{equation}
  \label{eq:comptonSCS}
  \frac{d  \sigma_{incoh}}{d\Omega}(\Vec{q}) =\frac{d\sigma_{KN}}{d\Omega}\frac{1}{F}  \int_{-\infty}^{+\infty}\!\! dt  j_X(\tau,t) S(\Vec{q},t),
\end{equation}
with 
\begin{equation}
  \label{eq:npisf}
  S(\Vec{q},t)=\sum_{j=1}^{N_a} s_j(\Vec{q},C_j(t)) \, ,
\end{equation}
and $s_j(\Vec{q},C_j(t))$ is the incoherent scattering function of the $j$-th atom/ion with electronic configuration $C_j(t)$.  Our method can also include the effect of the bandwidth of the XFEL pulse \cite{Ho-2020-NatComm} on the scattering response by convolving the differential cross section with the bandwidth profile.

Using the MC/MD method, we examine FIC and CDI of Ar$_{1415}$ and Ar$_{149171}$ clusters in an intense 5-keV pulse with a fluence of 3.5$\times$10$^{12}$ photons/$\mu$m$^2$, which corresponds to 10 times the fluence for saturating single ionization of Ar.  The choice of pulse fluence and photon energy is motivated by our previous work with a 2-fs pulse, in which these parameters can enable high {\Ka} and {\Kah} yield in a femtosecond temporal emission window.  Motivated by the capability to generate pulses with different pulse lengths, we examine the pulse duration dependence of FIC and CDI, including sub-fs pulses.  To explore the feasibility of imaging contrast with FIC, we investigate the scattering and fluorescence response for imaging dopant distribution in {\feo} with intense x-ray pulses.  Due to the small probability of fluorescence and Monte-Carlo nature of the calculation, a large number of replicas is needed to produce statistically converged fluorescence data.  For our calculations, we used 100 to 10,000 replicas such that the error in per atom fluorescence yield of the interested fluorescence channels is less than 0.1 \%.  A time step of 2 attoseconds is found to be sufficient to follow the electron processes and nuclear dynamics \cite{Ho-2020-PRA}.  For each calculation, we propagate the system dynamics for tens of femtoseconds as the fluorescense temporal emission extends beyond the pulse duration.

\begin{figure*}[t!]
\begin{center}
\includegraphics[width=0.95\linewidth]{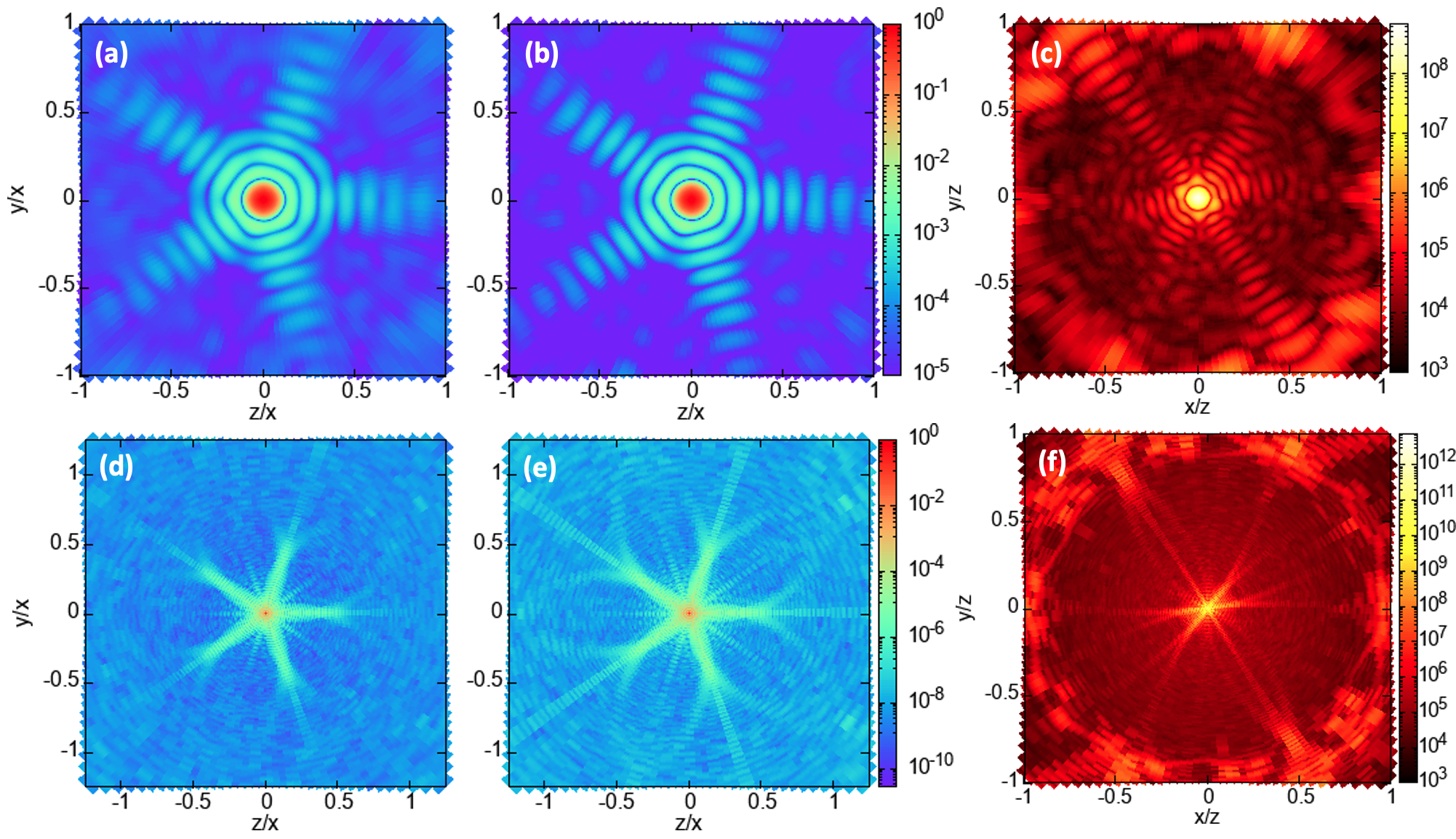}
\caption{Fluorescence intensity correlation and scattering patterns of Ar clusters in an intense x-ray pulse.  Panels (a), (b) and (c) are the FIC of {\Ka}, FIC of {\Kah} and the total differential scattering cross section  of Ar$_{1415}$, respectively, exposed to an 2-fs, 5-keV and 3.5$\times$10$^{12}$-photons/$\mu$m$^2$ pulse.  Panels (d), (e) and (f) in the bottom row are the same as those in the top row, except they are for Ar$_{149171}$. The fluence corresponds 10 times the single-ionization saturation fluence of Ar.   The geometry of the fluorescence and scattering detector is shown in Fig. \ref{fig:setup}. The FICs are computed with a fixed $\Vec{k}_1$,  which points along the -x axis, and $\Vec{k}_2$ scans across all detector pixels on the y-z plane.  In this geometry, the FIC images reveal the five-fold symmetry of the icosahedral structure of Ar$_{1415}$  and Ar$_{149171}$.}
\label{fig:scatteringvsfic}
\end{center}
\vspace{-20pt}
\end{figure*}


We note that from a given 2-D fluorescence intensity detection from an excited target in a fixed orientation, as shown in  Fig. \ref{fig:setup}, a large 3-D FIC data set can be gathered from all accessible pairs of $\Vec{k}_1$ and $\Vec{k}_2$ \cite{Classen-2017-PRL}.  This means one can gather the full 3D data set needed for 3D reconstruction by measuring fluorescence from a small number of sample orientations.  To illustrate the impact of intense-field XFEL on FIC, we focus on a subset of FIC data, in which $\Vec{k}_1$ is fixed along the -x axis and $\Vec{k}_2$ scans across all detector pixels on the y-z plane, as shown in Fig. \ref{fig:setup}.  The fluorescence detector is placed perpendicular to the XFEL propagation, where the probability of coherent scattering is suppressed.

\section{Results}

\begin{figure*}[t!]
\begin{center}
\includegraphics[width=1\linewidth]{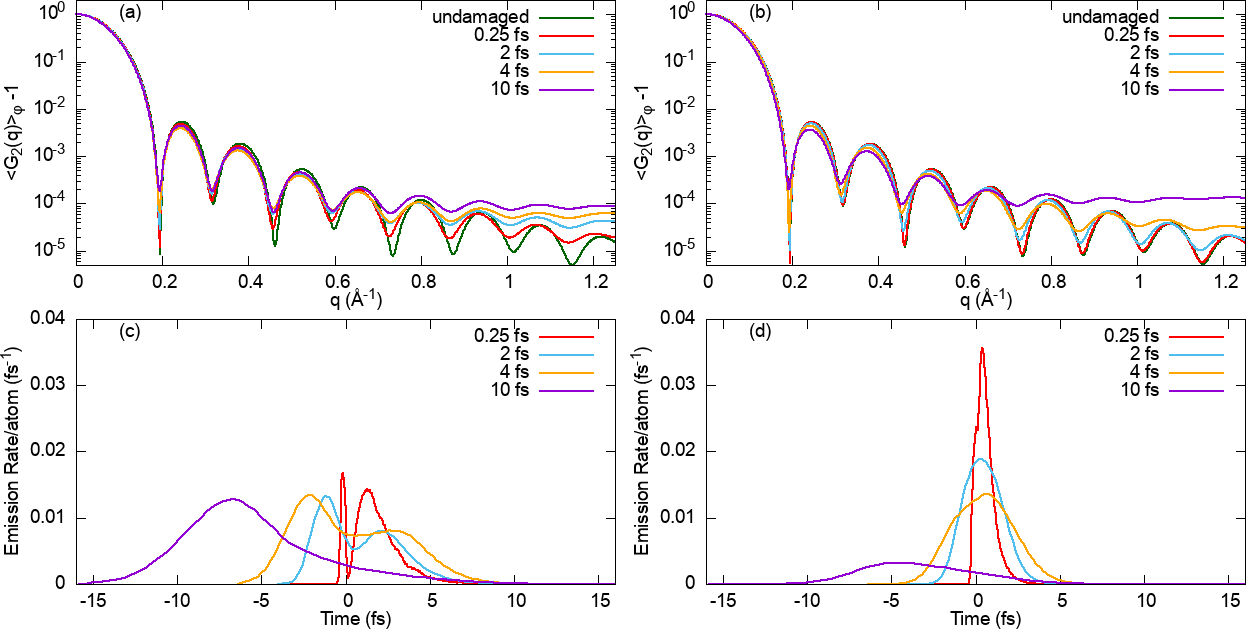}
\caption{\small Pulse duration dependence of FIC computed from (a) {\Ka} and (b) {\Kah} channels of Ar$_{1415}$ exposed to 5-keV, 3.5$\times$10$^{12}$-photons/$\mu$m$^2$ pulse.  For reference, the FICs computed from undamaged clusters are included. The bottom rows are the temporal emission profile of (c) {\Ka} and (d) {\Kah} channels.}
\label{fig:4panels}
\end{center}
\vspace{-20pt}
\end{figure*}

\subsection{Ar Clusters}

We begin our discussion by showing the fluorescence intensity and the FIC associated with the {\Ka} and {\Kah} channels of Ar$_{1415}$ and Ar$_{149171}$ subjected to an intense 2-fs pulse. Both Ar$_{1415}$ and Ar$_{149171}$ are constructed as icosahedral structures \cite{Ho-2020-PRA}. {\Ka} and {\Kah} are x-ray emission from ions with single and double vacancies in K shell respectively.  As expected, Fig. \ref{fig:setup} shows that the 2-d fluorescence intensity distribution, $\langle I(k)\rangle$, of Ar$_{1415}$ reveals featureless speckle pattern.  On average, each excited Ar$_{1415}$ cluster produces 0.0565 {\Ka} and 0.0572 {\Kah} photons per atom per pulse over a 4$\pi$ solid angle.  In comparison, the {\Ka} and {\Kah} yields found in an isolated Ar atom exposed to the same pulse are 0.048 and 0.040.  The higher {\Ka} and {\Kah} yields found in Ar$_{1415}$ are the result that recombination pathways, in addition to the direct photoionization pathway, are available in extended systems, as mentioned earlier.  Also, we point out the the similar yields between {\Ka} and {\Kah} channels are the results of nonlinear x-ray/matter interaction \cite{Ho-2020-PRA}.


Unlike the fluorescence intensity distribution, the computed FICs contain structural information as shown in Figs. \ref{fig:setup} and \ref{fig:scatteringvsfic}.  Despite having the same number of photons, FIC from the {\Kah} channel has a higher contrast than that of {\Ka}.  This is because {\Kah} are emitted over a narrow time window in which the atomic motion is limited, whereas {\Ka} spans a longer time emission window, which encompasses a larger atomic motion.  For larger systems, {\Kah} is more advantageous for structural determination than {\Ka} channel.  For example, Ar$_{149171}$, {\Kah} emission count is 1.5 times larger than that of {\Ka} emission and FIC of {\Kah} reveals higher contrast than that of {\Ka}, as shown in Figs. \ref{fig:scatteringvsfic} (d) and (e).  

Next, we show the pulse duration dependence of FICs computed from {\Ka} and {\Kah} of Ar$_{1415}$.  Figs. \ref{fig:4panels} (a) and (b) show the azimuthally averaged $G_2(q_f)$ for {\Ka} and {\Kah} emissions,
\begin{equation}
    \langle G_2(q_f) \rangle_{\varphi} = \frac{1}{2\pi} \int \! d \varphi \langle G{_2}(\Vec{k}_1,\Vec{k}_2,\{\Vec{R}_{f,j}\}) \rangle_{avg}
\end{equation}
where $q_f =|\Vec{k}_1-\Vec{k}_2|$.  For both of these emissions, we compute the FICs using 4 pulse durations of 0.25, 2, 4 and 10 fs.  The 0.25-fs calculations were motivated by the recently available XFEL pulses \cite{Huang-2012-PRL,Marinelli-2017-APL}.  For comparison, we include the FICs calculated without damage (i.e. the nuclei are assumed to be frozen).  The same pulse fluence of 3.5$\times$10$^{12}$ photons/$\mu$m$^2$ and photon energy of 5 keV were used in these calculations.  Fig \ref{fig:4panels} (a) shows that the degree of deviation of FIC of {\Ka} from that of the undamaged case increases with pulse duration.  Also, the degree of deviation increases with q, which is inversely proportional to the resolution in real space. Our results suggest that a shorter pulse will enable higher resolution imaging with higher contrast.

The examination of the pulse duration dependence of FIC from {\Kah} emission, as shown in Fig \ref{fig:4panels} (b), reveals the same trend found in {\Ka} emission.  However, the FICs of {\Kah} are less sensitive to radiation damage.  For example, the q-dependence FIC of {\Kah} at 0.25-fs is nearly identical to that of the undamaged case, whereas the q-dependence FIC of {\Ka} at 0.25-fs deviates from the undamaged profile starting at q > 0.4 {\AA} already.  Their difference is the result that the {\Kah} temporal emission profile is different from that of {\Ka}.  Fig \ref{fig:4panels} (c) and (d) show that for our pulses with a pulse duration of 0.25, 2 and 4 fs, the resulting temporal profile of {\Ka} is broader than that of {\Kah}.  These temporal profiles of {\Ka} shows a double-peak structure due to the presence of two pathways: the photoionization pathway and recombination pathway. The contribution of the photoionization and recombination pathway peak before and after t = 0, which is the peak of the XFEL pulses. On the other hand, these temporal profiles of {\Kah} remain a single-peak structure due to the contribution of the photoionization and recombination have significant temporal overlap \cite{Ho-2020-PRA}.

Fig \ref{fig:4panels} further shows that the fluorescence dynamics in 10-fs pulse is different from that in shorter pulse durations.  In particular, the temporal profile of {\Ka} is a single-peak distribution, indicating that the {\Ka} production is predominantly via the photoionization pathway.  The negligible role of the recombination pathway is due to substantial structural damage and expansion occurring before the peak of the pulse, leading to a reduced probability of electron-ion recombination.  Fig. \ref{fig:4panels} reveals the trend that the contribution of recombination pathway increasingly diminished in a longer (less intense) pulse.  Also, in a 10-fs pulse, the {\Kah} yield is about a factor 4 smaller than the {\Ka} yield.  The relatively small {\Kah} yield is due to it is more probable for a single-core-hole state to undergo inner-shell decay (Auger decay or emitting {\Ka}) than photoionization to produce a double core-hole state in a 10-fs pulse.

Our analysis shows that there is a trade off between high-resolution FIC images and the fluorescence intensity obtained with an intense pulse.  As pointed out earlier, a shorter pulse (0.25- vs 2-fs pulse) produces higher resolution and spectral contrast FIC images, but it also leads to a lower fluorescence intensity per pulse, as shown in Table \ref{table:fcount}.  For experimental considerations, FIC imaging with few-fs pulses might be experimentally more efficient than with sub-fs pulses since the latter tend to come with a lower number of photons.

\begin{table}[h]
\caption{The number of {\Ka} and {\Kah} photons/atom produced in Ar$_{1415}$ exposed to a 5-keV and 3.5$\times$10$^{12}$-photons/$\mu$m$^2$ pulse as a function of pulse duration.}
\label{table:fcount}
\begin{tabular}{  m{10em}  m{2cm} m{2cm}  } 
\hline
Pulse Duration (fs) & {\Ka} & {\Kah} \\
 \hline
0.25 & 0.0424 & 0.0431 \\ 
2 & 0.0565 & 0.0572 \\ 
4 & 0.0883 & 0.0622 \\ 
10 & 0.0827 & 0.0212 \\ 
\hline
\end{tabular}
\end{table}

\begin{figure}[t!]
\begin{center}
\includegraphics[width=1\linewidth]{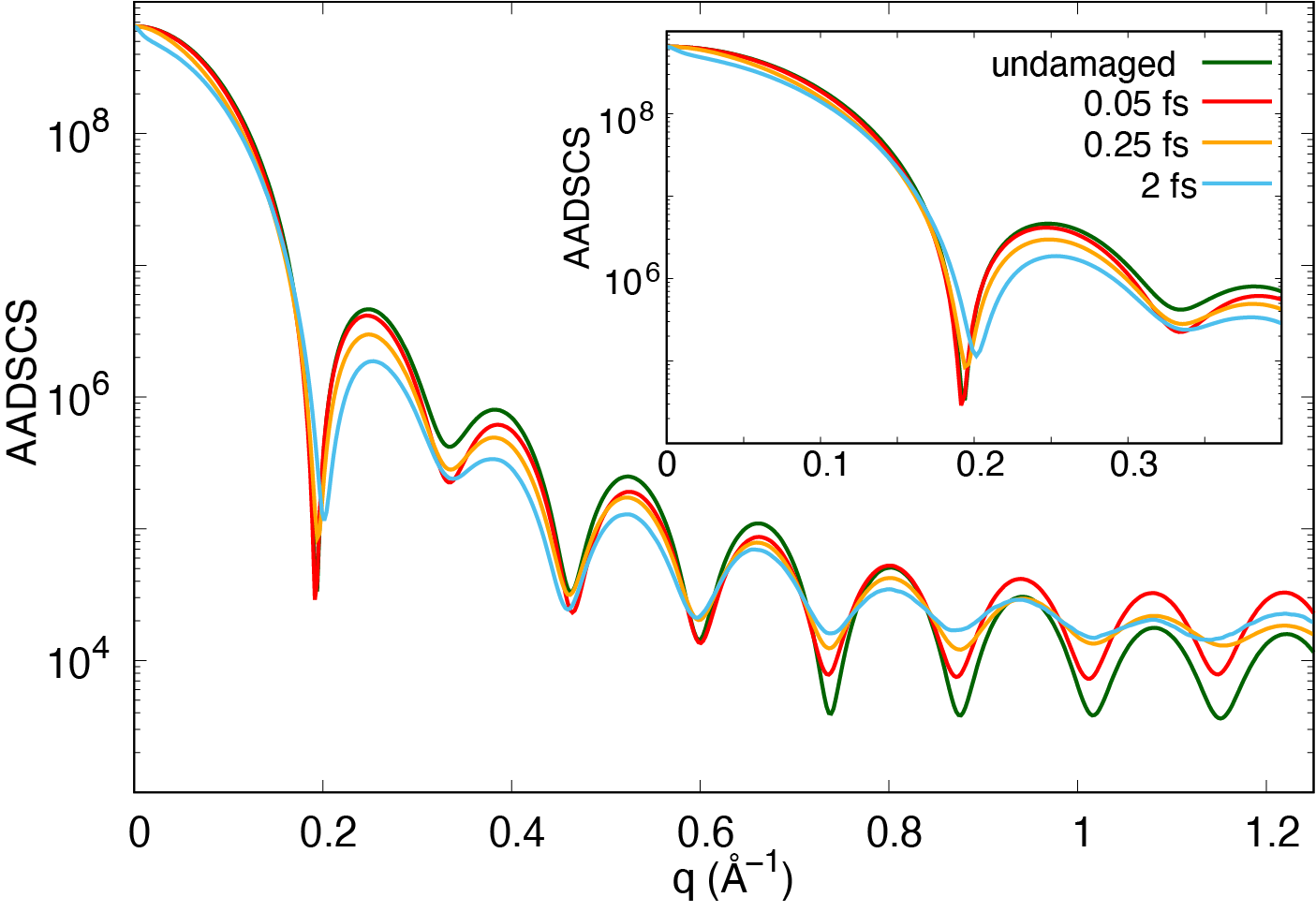}
\caption{\small Pulse duration dependence of azimuthally averaged differential scattering cross section (AADSCS) of Ar$_{1415}$ exposed to intense 5-keV x-ray pulses with a fluence of 3.5$\times$10$^{12}$ photons/$\mu$m$^2$.  The inset shows that the AADSCS from a 2-fs pulse deviates from that of the undamaged sample already at a small q. }
\label{fig:xdp}
\end{center}
\vspace{-20pt}
\end{figure}

Unlike the FIC, the quality of the scattering pattern benefits substantially from sub-femtosecond pulses in comparison to that from femtosecond pulses.  Fig. \ref{fig:xdp} plots the pulse duration dependence of the azimuthally average differential scattering cross sections (AADSCS),
$ \frac{1}{2\pi} \langle \frac{d  \sigma_{total}}{d\Omega}(\boldsymbol{q}) \rangle_\varphi$,
in units of classical electron radius square.  The scattering pattern from a 2-fs pulse deviates substantially from that computed by assuming no electronic and structural damage.  In particular, their difference starts immediately after the forward scattering region and the location of the first minimal shifts to a larger q.  This shift is a result that ultrafast ionization leads to significant distortion in the electron distribution of Ar clusters.  In contrast, using the same 2-fs pulse, the computed {\Ka} and {\Kah} FICs, which include the effect of atomic motion over a time window larger than the XFEL pulse, deviate from the corresponding FIC of the undamaged cases only at high q.  This means that the degree of atomic motion during a 2-fs pulse is not significant, and this further suggests that the expanding delocalized electron cloud is the main factor that reduces the quality of the scattering image in a 2-fs pulse. To further improve the quality of the scattering images of the Ar cluster, one can employ sub-fs pulses.  As shown in Fig. \ref{fig:xdp}, the degree of radiation damage, i.e. the extent of the expansion of the electron clouds and the nuclear motion, in 0.25-fs and 0.05-fs pulses is small. With reduced damage, the scattering efficiency per pulse and scattering cross section will also increase.

Also, our calculations also show that about 80 photons are scattered from Ar$_{1415}$ exposed to a 5-keV, 2-fs, 3.5$\times$10$^{12}$-photons/$\mu$m$^2$ pulse.  About 50 of these photons are scattered in the forward direction and in a small q region (q<0.2 {\AA}).  In comparison, about 80 {\Ka} and 81 {\Kah} photons are emitted isotropically from the cluster (not including emission from other fluorescence channels).  This suggests that, already in the small q region, the scattering signals and fringes will be washed out by the fluorescence photons.  
From the perspective of imaging small particles (molecules), fluorescence can be a dominant noise in the scattering pattern.  One can increase the scattering intensity with a higher fluence pulse, but there is a trade off between the higher scattering signal and quality of the scattering pattern.  For larger systems, the impact of fluorescence on scattering images will be less severe and it will be limited to a large q region as the scattering signal (total cross section) scales with $N_{a}^{4/3}$ (as discussed in work by Kirz and coworkers \cite{Kirz-1995-QRB}), whereas the fluorescence signal scales with $N_{emitter}$.


\begin{figure*}[t]
\begin{center}
\includegraphics[width=6.8in]{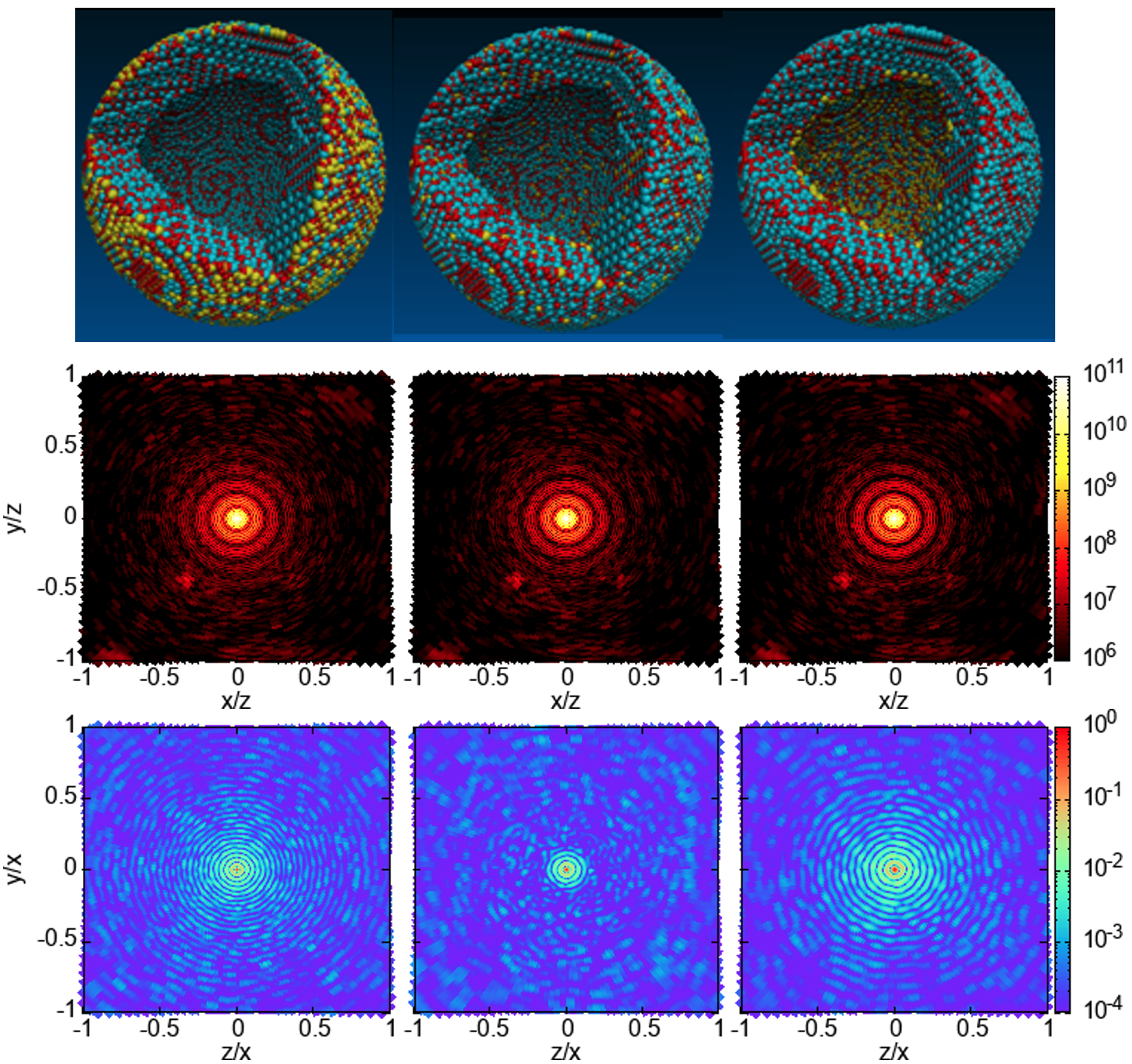}
\caption{Top panel: Structures of hollow core-shell {\feo} nanoparticles with Mo-dopants distributed randomly in the outer layer (left) and everywhere (middle) and inner layer (right) in the nanoparticle.  The yellow, cyan and red dots are molybdenum, iron and oxygen atoms. The panels in the second row are the corresponding differential scattering cross sections computed from an intense 2-fs, 5-keV XFEL, 6$\times$10$^{11}$-photons/$\mu$m$^2$ pulse in units of classical electron radius squared. The fluence is 5 times the single-ionization saturation fluence of Mo atom, but it is below the saturation fluence of O and Fe. The spot near (-0.4, -0.5) in each pattern is a Bragg peak, which is sensitive to the orientation of the cluster with respect to the x-ray propagation. The panels in the third row show the FICs, given by G$^{2}$-1 computed from {\La} fluorescence channel of Mo atom.  The same pulse parameter set is used for the FICs.  The scattering detector is perpendicular to the fluorescence detector, and their geometry is shown in Fig. \ref{fig:setup}.}
\label{fig:ironoxide2d}
\end{center}
\vspace{-20pt}
\end{figure*}

\begin{figure}[t!]
\begin{center}
\includegraphics[width=3.2in]{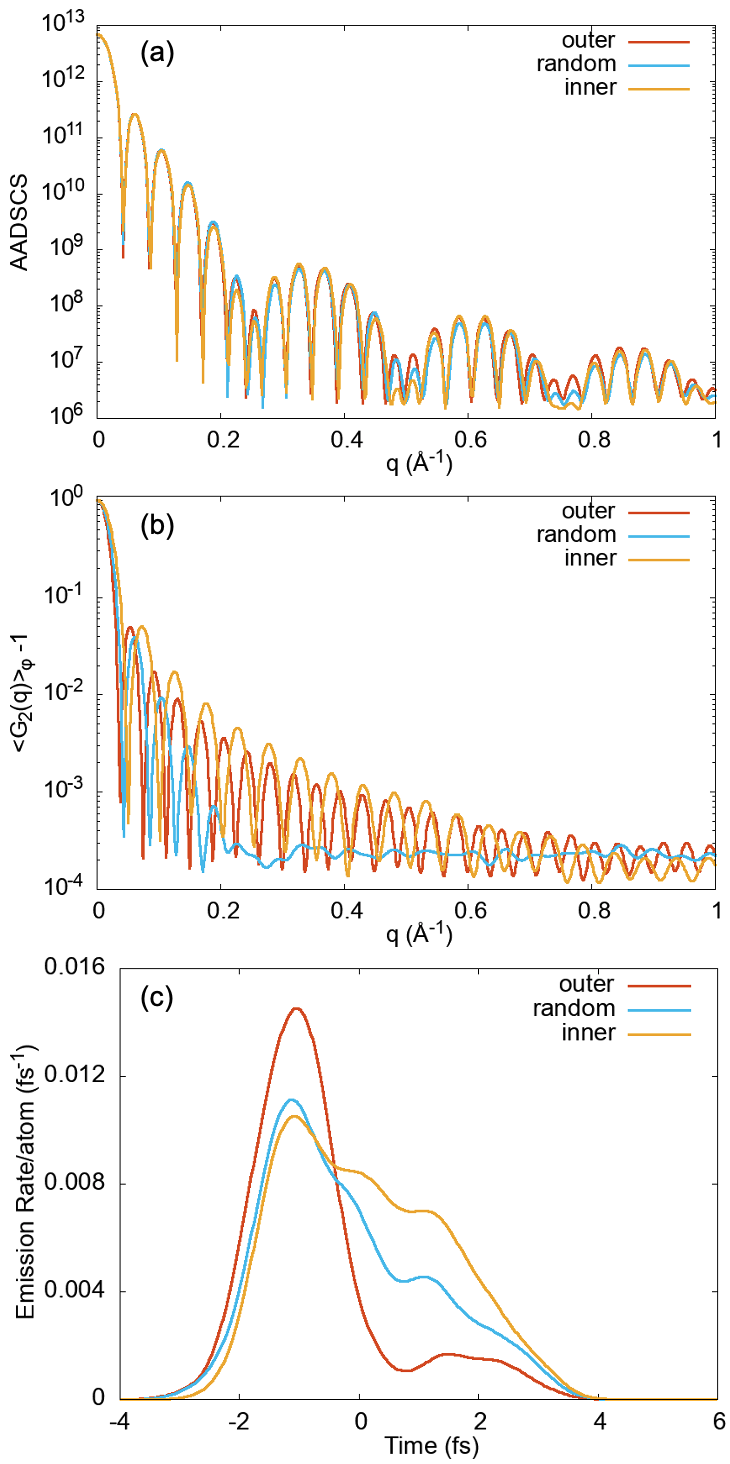}
\caption{(a) Azimuthally averaged differential scattering cross section (AADSCS) and (b) azimuthally averaged G$_2$(q)-1  as a function of momentum transfer, q, of the three iron oxide NPs with different dopant distributions shown in Fig. \ref{fig:ironoxide2d}. (c) Temporal emission profiles of {\La} emission. The same pulse parameters of 2 fs, 5 keV XFEL, 6$\times$10$^{11}$ photons/$\mu$m$^2$ were used.}
\label{fig:ironoxide1d}
\end{center}
\vspace{-20pt}
\end{figure}

\subsection{Doped Iron Oxide Nanoparticle}

%
%

Next, we explore CXD and FIC for imaging of elemental contrast in heterogeneous samples.  Our hollow core-shell structure of $\gamma$-Fe$_2$O$_3$ has a polycrystalline structure with an inner radius and outer radius of 6 and 8.5 nm.  We considered three different dopant distributions, as shown in Fig. \ref{fig:ironoxide2d}.  In the first structure,  the Mo dopants are located on the outermost 0.2 nm (i.e., between the radii of 8.3 and 8.5 nm) of the nanoparticle.  In the second structure, the dopants are randomly distributed in the sample volume.  In the third structure, the Mo atoms are doped in the inner 0.4 nm layer (i.e. between the radii of 6.0 and 6.4 nm).  In all these three nanoparticles, the relative abundance of Mo atoms (the ratio of the number of Mo atoms and the sum of Mo and Fe atoms) is 6\%.  

Each of these NPs has about 65.1-k iron, 92.3-k oxygen and 4.1-k Mo atoms.  Together with the electrons, each MC/MD calculation tracks more than 2.7 million particles.  These calculations were performed on the high-performance computer, Mira, at ALCF.  We calculate the scattering pattern and fluorescence intensity correlation patterns using a 2-fs, 5-keV pulse.  The pulse fluence is chosen to be 5 times the saturation fluence of the single ionization of Mo atom.  At this fluence level of 6$\times$10$^{11}$ photons/$\mu$m$^2$ and photon energy, the probability of photoionization for iron and oxygen atoms are below saturation.  
The pulse photon energy is chosen such that it is far from the resonances of the ground state and excited state hidden resonances in Fe, Mo, and O, and it induces L-shell fluorescence emission.

Our calculation shows that there are only small differences among the scattering patterns of these three structures, indicating that the scattering signals are not very sensitive to the dopant distribution in Fig. \ref{fig:ironoxide2d}. 
These small differences are a result of the scattering signals being dominated by Fe and O atoms, in which their distributions are similar in these hollow core NPs.  We note that the scattering amplitude can be considered as the superposition of two out-of-phase scattering amplitudes from two spherical particles with radii of 8.5 and 6 nm.  This superposition gives a beat pattern in Fig. \ref{fig:ironoxide2d} (a). 


Unlike the scattering processes, the fluorescence emission is elemental specific and electronic transition specific.  Figs. \ref{fig:ironoxide2d} (c) and \ref{fig:ironoxide1d} (b) shows that FICs computed from the {\La} channel of Mo reveal distinct fingerprints for these NPs with different dopant distributions already in the small q region.
Here, the pulse parameter is the same pulse parameter used in the scattering calculations.  In comparison, the averaged number of {\La} photons produced per pulse from each Mo atom are about 0.0351, 0.0332, and 0.0278 for inner-doped, randomly doped, and outer-doped structures.  These different yields are the result of the temporal emission profile and the relative contribution of the two {\La} pathways (direct photoionization and recombination pathways) depending on the location of the Mo atoms within the NP.  In general, for the Mo atoms residing near the surface layers of the NP, the {\La} events are mostly via the direct photoionization pathway and they take place earlier in the pulse (before t = 0, as shown in Fig. \ref{fig:ironoxide1d}(c)).  However, for Mo atoms residing deeper in the NP, they have a higher chance of undergoing {\La} via the recombination pathway in addition to the photoionization pathway, and they have a broader temporal emission profile.


\section{Summary}

In summary, we investigated FIC and CDI for imaging high-resolution structure of nanosized, non-periodic particles.  We showed that the parameters of intense XFEL pulse, which can enable many fluorescence channels, can serve as control knobs to optimize the fluorescence dynamics for FIC imaging.  
Using Ar clusters as prototypical systems, we investigated the pulse duration dependence of the FIC approach from {\Ka} and {\Kah} channels.  We show that, in few-fs or sub-fs pulses, the resulting FICs from {\Kah} in general reveal a higher contrast than those from {\Ka} due to {\Kah} having narrower temporal emission profiles. 
In comparison to the result from the 2-fs pulse, FIC from an 0.25-fs pulse offers only a slightly higher degree of contrast, but the total fluorescence intensity count is lowered by a factor of 2, suggesting that few-fs pulses may be sufficient for FIC measurements. 
On the other hand, it is more advantageous to perform scattering experiments with attosecond pulse than with 2-fs pulse as the short pulse reduces the sample electronic and structural damage, which leads to a higher scattering cross section and higher quality of scattering signals. Due to the isotropic nature of fluorescence emission, it can become a dominant noise in scattering patterns.

More interestingly, we presented the FIC of heterogeneous nanoparticle {\feo} structures with three different Mo dopant distributions in an intense x-ray pulse. We show that, while the scattering signals are dominated by the Fe and O atoms and reveal small differences between the three structures, the FICs from {\La} of Mo illuminates the three dopant distributions.  Our work suggests that FIC can be exploited for imaging the distribution of trace elements.  In the future, K-shell excitation in Mo, which requires 20-keV photons, will also be considered since XFEL pulses with photon energies up to 25 keV will be accessible at the EuXFEL. 

We point out that the current FIC analysis makes use of a subset of 2-point correlation functions.  For future work, it will be useful to exploit the full volumetric data from a fixed  sample orientation for structure analysis and determine the number of sample orientations needed for 3-D structural reconstruction.  Also, it will be interesting to explore higher-order coherence, G$^{(N)}$($\Vec{k}_1$,..., $\Vec{k}_N$), \cite{Classen-2016-PRL,Thiel-2007-PRL} derived from the x-ray fluorescence of a collection of atoms exposed to intense XFEL pulses.  


\section{Acknowledgement}

We thank Carsten Fortmann-Grote for the early version of the data files containing the atomic positions of the three Mo-doped iron oxide NPs.  We also thank Elena Shevchenko for the fruitful discussion about the atomic structure and synthesis of the Mo-doped iron oxide NPs.
This material is based on work supported by the U.S. Department of Energy, Office of Basic Energy Sciences, Division of Chemical Sciences, Geosciences, and Biosciences through Argonne National Laboratory. Argonne is a U.S. Department of Energy laboratory managed by UChicago Argonne, LLC, under contract DE-AC02-06CH11357. This research used resources of the Argonne Leadership Computing Facility at Argonne National Laboratory, which is supported by the Office of Science of the U.S. Department of Energy, Office of Science, under contract number DE-AC02-06CH11357.

\section{Data Availability}
The data that support the findings of this study are available
from the corresponding authors upon reasonable request.

\bibliography{PJHbib}

\end{document}